\documentclass[conference]{IEEEtran}
\IEEEoverridecommandlockouts
\usepackage{cite}
\usepackage{amsmath,amssymb,amsfonts}
\usepackage{algorithmic}
\usepackage{graphicx}
\usepackage{textcomp}
\usepackage{xcolor}
\usepackage{fancyhdr}
\def\BibTeX{{\rm B\kern-.05em{\sc i\kern-.025em b}\kern-.08em
    T\kern-.1667em\lower.7ex\hbox{E}\kern-.125emX}}
\begin{document}

\title{Cognitive Load and Productivity Implications in Human-Chatbot Interaction
}

\author{\IEEEauthorblockN{Johanna Schmidhuber}
\IEEEauthorblockA{\textit{Management, Communication \& IT} \\
\textit{MCI - The Entrepreneurial University}\\
Innsbruck, Austria \\
sj8377@mci4me.at}
\and
\IEEEauthorblockN{Stephan Schl\"{o}gl}
\IEEEauthorblockA{\textit{Management, Communication \& IT} \\
\textit{MCI - The Entrepreneurial University}\\
Innsbruck, Austria \\
stephan.schloegl@mci.edu}
\and
\IEEEauthorblockN{Christian Ploder}
\IEEEauthorblockA{\textit{Management, Communication \& IT} \\
\textit{MCI - The Entrepreneurial University}\\
Innsbruck, Austria \\
christian.ploder@mci.edu}

}

\maketitle

\thispagestyle{fancy}
\lfoot{Preprint}
\cfoot{}
\rfoot{}

\begin{abstract}
The increasing progress in artificial intelligence and respective machine learning technology has fostered the proliferation of chatbots to the point where today they are being embedded into various human-technology interaction tasks. In enterprise contexts, the use of chatbots seeks to reduce labor costs and consequently increase productivity. For simple, repetitive customer service tasks such already proves beneficial, yet more complex collaborative knowledge work seems to require a better understanding of how the technology may best be integrated. Particularly, the additional mental burden which accompanies the use of these natural language based artificial assistants, often remains overlooked. To this end, cognitive load theory implies that unnecessary use of technology can induce additional extrinsic load and thus may have a contrary effect on users' productivity. The research presented in this paper thus reports on a study assessing cognitive load and productivity implications of human chatbot interaction in a realistic enterprise setting. A/B testing software-only vs.  software + chatbot interaction, and the NASA TLX were used to evaluate and compare the cognitive load of two user groups. Results show that chatbot users experienced less cognitive load and were more productive than software-only users. Furthermore, they show lower frustration levels and better overall performance (i.e, task quality) despite their slightly longer average task completion time.
\end{abstract}


\section{Introduction}
Chatbots have been at the heart of computer science research ever since Alan Turing proposed his famous test on evaluating machine intelligence~\cite{machinery1950computing}. But, it is only since respective systems started to learn, reason and make decisions, that companies have put their hopes up as to the opportunities said technology may offer~\cite{mattioli2019gestion}. 
Improvements have been particularly visible in the fields of information retrieval and process automation, leading to a substantial uptake in chatbot use over the last years~\cite{czarnecki2019rolle}. Yet, chatbots are no humans, and users often have little to no experience in communicating with them. Thus, respective interactions may require significantly more cognitive effort. To this end, the goal of the research presented in this paper was to focus on mental workload in human-chatbot interaction. In doing so, we investigated the use of a B2B software product and therein compared chatbot-based task execution (i.e., text in- and output) with traditional software-based task execution, measuring users' perceptions of cognitive load as well as their productivity. These investigations were guided by the following two research questions:

\begin{enumerate}
    \item \textit{In which way does the use of a chatbot affect users' mental effort when interacting with a new software product?}
    \item \textit{To what extend does the use of a chatbot affect users' productivity when interacting with a new software product?}
\end{enumerate}

\noindent
The following report of these analyses starts with Section~\ref{sec:relwork}, discussing relevant related work. Next, Section~\ref{sec:methodology} elaborates on our research methodology, Section~\ref{sec:results} presents our results and Section~\ref{sec:discussion} discusses their impact. Finally, Section~\ref{sec:conclusion} concludes with some limitations attached to our findings and proposes directions for future research. 

\section{Definitions and Related Work}\label{sec:relwork}
Derived from Mauldin's term \emph{chatterbot}~\cite{mauldin1994chatterbots}, more recent literature holds different but mostly related definitions for text-based conversational digital assistants. Czarnecki et al.~\cite[p. 801]{czarnecki2019rolle}, for example, describe chatbots as autonomous software agents whose main characteristic is the ability to interact with humans via natural language. Dale defines them as \textit{``any software application that engages in a dialog with a human using natural language''}~\cite[p. 813]{dale2000symbolic}. Brandtzaeg \& Folstad call them \textit{``machine agents serving as natural language user interfaces for data and service providers''}~\cite[p. 377]{brandtzaeg2017people}, and  Mayo~\cite{mayo2017programming} adds that chatbots are often provided on messaging platforms, use some kind of intelligence, and require a user interface so as to communicate with a human. 



Throughout the last 25 years we have seen significant improvements in chatbot capabilities -- improvements which led to numerous fields of application.

\subsection{Chatbot Fields of Application}
Recent years have seen chatbots propagating into many different fields of application, including but not limited to the production and service industry, education, healthcare, entertainment, tourism, culture, and finance~\cite{io2017chatbots}. Customer service management, e.g., has become well-known for its chatbot applications~\cite{dubois2019chatbot}, but also end-to-end business processes have increasingly been automated by chatbots, as they can provide complex services while keeping staff costs low~\cite{ayachitula2007service,kalia2017quark}. Service desk chatbots, e.g., increase efficiency in issue management~\cite{espig2019bewertung}, and text- and voice-based digital assistants support customer relationship management~\cite{daisy2020stimulus}. 

From an educational perspective, chatbots assist in learning foreign languages~\cite{dokukina2020rise,zadrozny2000natural}, complement lecturers' teaching activities and support student administration~\cite{medrano2019empleo,hien2018intelligent}.

In the travel industry, one finds chatbots being used for flight management as well reservation and transportation management (an area which has seen a significant chatbot uptake during the last year as the travel industry has suddenly been facing countless customer calls, cancellation and refund requests due to the COVID-19 pandemic)\footnote{Online: https://apex.aero/articles/airlines-chatbots-automate-customer-service-requests-soar/ [accessed: February 24\textsuperscript{th} 2021]}. 


In healthcare, chatbots are taking over patient admission tasks in hospitals and provide domestic support when coping with chronic illnesses. Here, developments go as far as that medical experts in France investigated the question whether patients would communicate more openly with a neutral, non-human chatbot than with their physician, pharmacist, or nurse~\cite{chaix2018impact}. Medical chatbots furthermore help physicians with complex tasks such as patient diagnostics. To this end, Battineni et al.~\cite{battineni2020ai} developed a chatbot concept for diagnosing COVID-19 cases, evaluating the severity of symptoms, and recommending countermeasures to infected patients. 

Finally, also the public sector seems to benefit from chatbot use. Gros Salvat et al.~\cite{gros2020codiseno}, e.g., used a chatbot to help  migrants in Barcelona who had little experience interacting with mobile technologies and consequently struggled to use them for more complex tasks. 

\subsection{Chatbots in Business Settings}
Although the above discussion shows that chatbots are increasingly used in the interaction with clients, we should highlight that supporting business workflows is an entirely different task. Underlining this, Stieglitz and colleagues define  
\textit{``An Enterprise Bot [to be] an automated user service that provides casual and conversational interactions with complex enterprise systems and processes. Enterprise Bots can, e.g., answer questions or perform smaller tasks. A user can interact with [them] by just typing or speaking a request in natural language. [They] can only act in a passive way, meaning [they] must actively be triggered by the user''}~\cite[p. 4]{stieglitz2018defining}. 


Unlike human workforce, chatbots are easily scalable, extending traditional 1:1 interactions between single employees to 1:n interactions between one chatbot and many employees. In other words, one single chatbot is capable of simultaneously processing multiple user requests~\cite{vincze2017virtual}, which reduces costs and consequently increases effectiveness~\cite{rahman2017programming}. Yet, added value is not solely related to financial benefits, as the increased user satisfaction chatbots often tigger~\cite{chung2018chatbot} may also impact on customer and/or employee retention~\cite{maas2019conversational,shaikh2019survey}. 

\subsection{Chatbots and Cognitive Load}
Cognition describes the human processing of incoming information~\cite[p. 119]{cohen2004voice}, while Cognitive Load (CL) addresses the question of how this respective information is buffered by the brain's limited storage capacity. It is considered a complex, multidimensional concept~\cite{hart1990workload} whose study is grounded in the Cognitive Load Theory (CLT)~\cite{sweller1988cognitive}. Cooper describes CL as the \textit{``total amount of mental activity imposed on working memory at an instance in time''}~\cite{cooper1998research}. In literature, CL is better known as Mental Workload (MWL)~\cite{orru2018evolution}, where, according to Hart~\cite{hart2006nasa}, it is the cost for humans to accomplish the requirements of a mission, including factors such as stress, accidents, fatigue, or illness. 

Research indicates that high CL can impair task performance~\cite{yurko2010higher}, where task performance is based on two factors, i.e., (1) task complexity, including cognitive and physical workload, and (2) humans' individual capabilities, such as skills or experiences~\cite{kartali2019real}. While Miller's earlier work on cognitive load~\cite{miller1956magical} showed, that `cognitive chunking' may help in dealing with complex tasks, CLT suggests avoiding unnecessary or wrongly used technology, for such would increase cognitive and/or physical workload, and consequently decrease productivity. 



\subsection{Chatbots and Productivity}
The concept of productivity, initially used by Quesnay~\cite{quesnay1766analyse} in agriculture, is a measure of success, generally defined as: 

\begin{equation} \label{eq:2}
P=\frac{Output}{Input}
\end{equation}


CL is considered one of the productivity input factors, where low CL usually increases productivity by shortening the task completion time~\cite{kc2019task}. Conversely, King \& Ehrenberg consider high CL a ``\emph{subtle productivity vampire}"~\cite{king2020productivity}. Also, Fried et al.~\cite{fried1993measurement} show a strong link between productivity and technology; although this relationship seems rather ambiguous, for according to them the use of technology can enhance productivity but may as well cause impairment~\cite{fried1993measurement}. To this end, Karr-Wisniewski \& Lu found three factors for technology-caused loss of productivity, i.e., communication, information, and system feature overload~\cite{karr2010more}. Such is particularly worrying, as our society seems to be increasingly burdened by information overload, primarily caused by multi-tasking and the use of ever more technologies~\cite{byyny2016pharos}. Furthermore, it was shown that extraneous cognitive load caused by complicated computer systems frustrates professionals (e.g., physicians) and impairs their performance/productivity. Intuitive, `natural' user interfaces, on the other hand, help reduce CL and therefore may enhance productivity~\cite{king2020productivity}. Hence, although the use of chatbots may reduce users' CL in interactive tasks~\cite{amiot2020trustworthy}, due to their naturalness, it is to be explored  whether such also increases, or potentially hampers, productivity. 





\section{Methodology}\label{sec:methodology}
In order to investigate and compare cognitive load and respective productivity impacts related to (A) indirect interactions with a software product via chatbot, and (B) direct interactions with a software (i.e., sans chatbot), we used NASA's Task Load Index (TLX) research model~\cite{hart1986nasa}. Originally developed for aerospace applications, the NASA TLX belongs to a set of subjective measures according to the CLT\cite{orru2019evolution}, and as such is still considered one of the most established CL measures to date~\cite{dearing2018assessing,hart2006nasa}. It aims to determine a person's subjectively perceived MWL right after completing a given task. Evaluation is split into two parts, i.e., rating and weighting. Six rating factors, each of which is measured on a 20-point sub-scale, aim to evaluate the felt experience. Since the perceived relevance of each factor varies from person to person, study participants are furthermore asked to rate each factor's task relevance. That is, factors are presented as 15 individual factor pairs, and participants are then asked with each pair to choose the one factor that seems more relevant for the given task. Although today many studies exclude the weighting of these factors in their analysis, Hart emphasizes that it makes the NASA TLX significantly more sensitive and/or accurate and thus should not be ignored~\cite{hart2006nasa}.

\subsection{Uses Cases and Tasks}
Using the two above outlined interaction modes, i.e., (A) and (B), we aimed to compare the level of cognitive load users must bear when interacting with a B2B software that is unknown to them. Data was collected via A/B testing the use of said software with and without a text-based MS Teams chatbot.
In order to design realistic use cases we first deployed a survey to all of the 100 companies which currently use the chosen B2B software. Although we only received 19 responses, such helped identify valid scenarios as well as existing interaction difficulties for the following 4 experimental use cases:
\begin{enumerate}
    \item Retrieval of and direct access to data and reports;
    \item Provision of software information;
    \item Access to user documentation and help;
    \item General user questions;
\end{enumerate}

\noindent
From this we derived the following 7 scenarios, containing a total of 13 tasks to be completed:
\begin{itemize}
  \item Scenario 1: (1) Please find out what type of software  \textit{X} is, and (2) for what size of company it is suitable for (no concrete number required). (3) Which version of \textit{X} is running?
    \item Scenario 2: Find all reports with the name ``Organisational Structure''. (4) How many do exist?
	(5) How many reports exist for ``System''?
	\item Scenario 3: Search for ``Bike''. (6) How much data do you find for this search term? (7) Did you find reports, elements, or both for this search term?
	\item Scenario 4: Search for the product ``Mountainbike''. (8) Find potential manufacturers.
	\item Scenario 5: (9) Does the IT service ``Second-Level-Support'' exist? (10) What else do you find for ``Support''?
	\item Scenario 6: (11) Use the documentation to find the answer to what the name of the software stands for. 
	\item Scenario 7: Search for the ``Technologies'' diagram. (12) What is its full name? (13) What colors does the diagram have?
\end{itemize}



\subsection{Chatbot}
The chatbot was realized as an Azure-hosted JavaScript web application based on Microsoft's Bot Framework\footnote{Online: https://dev.botframework.com/[accessed: January 06\textsuperscript{th} 2021]}~\cite{mayo2017programming}, and integrated into Microsoft Teams via App Studio. In particular, we used Microsoft's Cognitive Services LUIS\footnote{Online: https://www.luis.ai/[accessed: January 15\textsuperscript{th} 2021]} to identify and score the intents of user messages, and the QnA Maker API\footnote{Online: https://www.qnamaker.ai/[accessed: January 16\textsuperscript{th} 2021)} \cite{shaikh2019developing} to generate predefined question-answer pairs based on a given knowledge base. Consequently, the chatbot queried the software's data elements and reports, directed users to its documentation or corporate website, returned predefined answers from the QnA knowledge base, or asked for additional input in cases where it could not identify the given intent.

\subsection{Participants}
Given the existing COVID-19 restrictions, we followed a hybrid approach to A/B testing. That is, $13$ of the $n=22$ experiment participants conducted the testing in a test laboratory on-site. The others tested remotely but confirmed that they had not been interrupted during the experiment. Recruited participants included students, employees as well as CEOs of several SMEs covering the fields of logistics, IT, tourism, and finance.
We used a convenience sampling strategy which aimed at representing diversity in age, gender, software skills, and experience in professional business software use. Also, the participants' living/working distance from the test laboratory was considered so as to allow for (partial) on-site testing. While they had varying expertise with respect to software use in general and different professional backgrounds, none of them had any prior experience with the tested B2B software. Participation was entirely voluntary and in accordance with European data protection regulations (i.e., GDPR) including respective  informed consent procedures. 


\subsection{A/B Test Setting and Procedure}
Our empirical analysis took place in August 2020 and was previously approved by the school's Research Ethics board.
For the experimental test setting, we randomly split participants into two equally sized groups, i.e., (A) chatbot and (B) software-only. 
A separate pre-test with one chatbot user and one software-only user was conducted to ensure the validity of the test setting~\cite{Hu2014}. 
Participants of Group A could use the chatbot to help tackle the above defined set of tasks, while those of Group B were not given this additional support feature. 
As for the subsequently completed NASA TLX, cognitive load evaluation was divided into two parts: (1) rating and then (2) weighting the six factors which described the experience felt by the participants directly after task completion (i.e., $Mental Demand=MD$, $Physical Demand=PD$, $Temporal Demand=TD$, $Effort=EF$, $Frustration=FR$ and $Performance=PER$)~\cite{hart1986nasa}). Thus, in both groups, all participants first individually evaluated these six factors on a 20-point scale~\cite{hart1986nasa} after each task they completed. After the experiment, they then weighted the six factors by comparing 15 factor pairs. 

We followed a strict between-group comparison approach~\cite{downey2007group}, for which neither participants of Group A nor those of Group B were aware of the other setting or the alternative form of interaction. All participants completed all tasks, resulting in a complete data set of $n=22$ measurements. 

\subsection{Data Cleansing}
The rated and weighted results of the NASA TLX created the basis for the CL analysis. Non-numerical data required coding into numerical values, whereas Yes/No answers were coded into binaries, i.e., $1=Yes$ and $0=No$. The 15 factor comparisons were converted into $1=selected\:factor$ and $0=not\:selected\:factor$. The computation of the factor scores and consequently the overall NASA TLX score per group required (1) the calculation of the six factor weights per participant, (2) the factor-weighted individual ratings per task, (3) the raw rating mean per participant, (4) the factor-weighted rating mean per participant, (5) the raw group score and finally (6) the factor-weighted group score (cf. Equation~\ref{eq:1} where rating factor $i=[0;6]$; group $G={A,B}$; group participant $p=[0;11]$). 
\begin{equation} \label{eq:1}
NASA\; TLX=\frac{1}{i*p}*\sum_{k=1}^{i}{rating_{weighted_{i,p,G}}}
\end{equation}

\noindent
Results were verified via the NASA TLX online calculator\footnote{Online: https://testscience.org/about-us/ [January 16\textsuperscript{th} 2021]}. 
As for task success, we coded participants' achievements as $1=correct$ and $0=false/no\; answer$, disregarding potential quality variations in participant performance.


 

\section{Results}\label{sec:results}
The collected NASA TLX data shows reliable internal consistency, i.e., Cronbach’s $\alpha$ = $0.797$ (Group A) and $0.938$ (Group B). 
Results show that, on average, software-only users (Group B) perceived cognitive load 32\% higher than chatbot users (Group A) perceived it ($28.6$ vs. $21.6$). 
Evaluating the single TLX factors we furthermore see, that while the weighted averages of $MD$, $TD$, $EF$, and $FR$ are lower for chatbot users, $PD$ and $PER$ are rated higher than those of software-only users. 
As for overall performance, in Group A, the 11 participants provided on average 86.2\% correct answers to the 13 questions, with $>90\%$ correct answers for questions (1), (2), (3), (4), (6), (9), (11), (12), and (13), $82\%$ for (10), $73\%$ for (8), $64\%$ for (7), and $55\%$ for (5). In comparison, only $63.8\%$ of Group B's answers to the 13 questions were correct, with $100\%$ correct answers for question (3), $82\%$ for (2), (9) and (11), $73\%$ for (6), (12) and (13), $64\%$ for (1), (4), (8), $36\%$ for (7) and (10), and finally $0\%$ for question (5). 

Focusing on $PD$, the analysis further showed that although chatbot users found the interaction physically more demanding than software-only users, the weight for this factor was rather low, both for Group A ($0.3$) and for Group B ($0.4$). The greatest difference in cognitive demand perception was found with respect to $FR$. Here, we see a clear tendency that chatbot users perceived the tasks less frustrating than software-only users. And finally, also with respect to $EF$, results point to a significant difference, indicating that chatbot users found tasks less demanding than software-only users.




Experiment completion times in Group A ranged from 22 to 45 minutes with an average of 33 minutes, while in Group B the fastest experiment completion took 17 minutes and the longest 50 minutes, averaging 29 minutes. Some users took two (Group A) or even three (Group B) times longer than others to complete the same tasks, either due to more problems with the tasks, more patience to find the right answer, or lower perceived time pressure.

Regarding users' productivity, we followed the rather simplified calculation proposed by Fried e al.~\cite{fried1993measurement} (cf. Equation~\ref{eq:2}). To this end, we considered the average number of correct answers given by a participant in the experiment as $Output$, and the average time to complete the experiment as well as the average amount of perceived CL (i.e, the NASA TLX score) as $Input$. 
Based on this input-output assumption, the relation between $P_{Group A}$ and $P_{Group B}$ computes to $\frac{P_{A}}{P_{B}}=1.60$. In other words, the average productivity of chatbot users was 60\% higher than that of software-only users.
    
   

\section{Discussion}\label{sec:discussion}
Our results show that chatbot users perceived task execution over three times more physically demanding than software-only users. Although such may easily be explained by comparing the physical effort needed to manually type-write text (i.e., to chat) with the effort required to click a mouse button or to type single words, it is still important to acknowledge users' awareness of this additional physical burden, and it should thus be considered in future design decisions concerning natural language based interaction (note: a text prediction feature, e.g., may help reduce typing and thus lower the physical effort).  

The relatively high frustration of software-only users compared to chatbot users, however, demands some more reflection. Both groups weighted the $FR$ most heavily of all factors, i.e., the level of frustration. As mentioned in literature, frustration seems to be an important factor in users' satisfaction\footnote{Online: https://www.accenture.com/t00010101T000000\_\_w\_\_/br-pt/\_acnmedia/PDF-45/Accenture-Chatbots-Customer-Service.pdf [accessed: January 15\textsuperscript{th} 2021]}~\cite{king2020productivity,roxin2019information}. Chatbot users invested more time but were less frustrated, suggesting that the chatbot positively impacted on the interaction experience. It always returned an answer, even if it did not understand users' intentions. Thus, users might have perceived it to be a helping assistant and therefore felt less frustrated when a search turned out to be unsuccessful. 
Software-only users, on the other hand, searched for information on their own, sometimes running into dead ends. Frustration may have arisen more quickly due to impatience and/or time pressure, which would also explain their higher $TD$ scores; or, alternatively it may have been rooted in some personal bias caused by past frustrating (first) interactions with software. To this end, the lack of experience participants had with chatbots might also explain the higher frustration tolerance with Group A (chatbot), as people often tend to be more patient when using rather novel and innovative technologies. 

As for productivity, results have shown that software-only users completed the experiment on average four minutes faster but rated $TD$ higher and $FR$ very high in comparison. Also, their $EF$ doubles the score of chatbot users. Possible explanations for these differences are (1) a higher personal bias of Group B due to past interaction with software, (2) the complexity of the used B2B software, (3) a higher level of perceived stress that caused impatience, and potentially faster task cancellation, or (4) difficulties in understanding the questions. 

Assuming we only consider the above mentioned input and output, the overall productivity of chatbot users in our experimental settings was 60\% higher than that of software-only users. Additionally, we found that chatbot users gathered more correct information. Thus, despite the small sample size and the rather simplified productivity calculation, results do point to performance improvements through chatbbot use -- even in cases were task completion times increased, as this also increased the number of correct answers. 
Interestingly, though, chatbot users evaluated their performance worse than software-only users. And while this finding may be subject to chance, for the $PER$ scores differ little in their absolute values, it indicates that chatbot use may be accompanied by a subjective underestimation of individual task performance.

Finally, our findings confirm previous studies in that information retrieval via chatbot is perceived to be less strenuous and overall simplifies human-technology interactions~\cite{valtolina2018chatbots,jiang2020response}. 
With respect to our research questions we may thus argue that, compared to software-only interaction, chatbot interaction (1) decreased users' perceived mental effort, and (2) increased their overall productivity.

\section{Conclusions and Potential Future Work}\label{sec:conclusion}
Today's chatbots are considered a valuable tool to support simple, frequent, repetitive tasks, both in private as well as in business contexts. For more complex use cases, however, the technology's value is less obvious as supervision, correction and overhead cost may be difficult to judge. All these may be considered additional physical and cognitive human user input (i.e, burden), and thus eventually effect the overall productivity of chatbot use. With our study we wanted to shed some light on these input variables and how they effect said productivity.

Concluding, we found that chatbot interaction may require less CL than software interaction, and furthermore that such potentially increases productivity. Although the quality and respective success of human-technology interaction in business settings depends on many more factors than those considered by this study, we believe these results do provide some relevant insights, which may also trigger further investigation.  

First, we found that information retrieval via chatbot yielded better, i.e. more correct, results. That is, software users were either less patient with finding the correct information or the efforts concerning this information retrieval task felt to be too high. This insight is particularly relevant, as the software we used as a test case in this experiment contained significantly less data than it would when used in a real customer setting. With more data, the likelihood of finding the right information would be even less, for which we believe that the chatbot may not only act as a task facilitator, but also as a `quality enhancer' and potential `frustration reducer'. 
Whether such would increase the productivity further, however, requires additional investigations using a greater sample size.


Second, although we found that the use of chatbots may have positive implications with respect to the correctness of information retrieval tasks and consequently impact positively on the productivity of knowledge workers, its implementation is often costly and the required effort for its continuous training is substantial. We used Microsoft's Bot Framework in order to integrate the chatbot into the MS Teams platform, thereby creating the illusion of it being a full-fledged `workmate'. Easier and less resource intensive implementations may, however, help with the proliferation of the technology. Also, we believe that the more chatbots are integrated into such collaborative tools, the higher their potential benefit. Yet, what remains unclear, is whether this also fosters chatbot acceptance, which may require respective investigations concerning long-term chatbot adoption. 

Finally, our study showed that several chatbot users were surprised about some social abilities of the technology. Particularly, with respect to its expressed kindness and active helpfulness. Whether such is simply a first impression or has the potential to be a long-lasting feeling regarding the interaction with AI technology, however, needs to be seen. After all, much of human-technology interaction is highly subjective. Other users may have had rather frustrating experiences with chatbots and thus feel very different about this technology. Hence, again, future work will need to focus on longitudinal studies of chatbot use and its implications.

\bibliography{ichms}
\bibliographystyle{IEEEtran}

\end{document}